\documentclass[oldversion,longauth]{aa} 
\usepackage{natbib}
\usepackage{txfonts}
\usepackage[english]{babel}
\usepackage{graphicx}
\usepackage{color}

\newcommand{\gsim}{\;\lower.6ex\hbox{$\sim$}\kern-7.75pt\raise.65ex\hbox{$>$}\;}
\newcommand{\lsim}{\;\lower.6ex\hbox{$\sim$}\kern-7.75pt\raise.65ex\hbox{$<$}\;}

\begin{document}
   \title{The VVDS type--1 AGN sample: The faint end of the luminosity function}


\author{A. Bongiorno \inst{1}
\and G. Zamorani \inst{2} 
\and I. Gavignaud \inst{3}
\and B. Marano \inst{1} 
\and S. Paltani \inst{4,5}
\and G. Mathez\inst{6}
\and J.P. Picat \inst{6}
\and M. Cirasuolo \inst{7}
\and F. Lamareille \inst{2,6}
\and D. Bottini \inst{8}
\and B. Garilli \inst{8}
\and V. Le Brun \inst{9}
\and O. Le F\`evre \inst{9}
\and D. Maccagni \inst{8}
\and R. Scaramella \inst{10,11}
\and M. Scodeggio \inst{8}
\and L. Tresse \inst{9}
\and G. Vettolani \inst{10}
\and A. Zanichelli \inst{10}
\and C. Adami \inst{9}
\and S. Arnouts \inst{9}
\and S. Bardelli \inst{2}
\and M. Bolzonella  \inst{2} 
\and A. Cappi    \inst{2}
\and S. Charlot \inst{12,13}
\and P. Ciliegi    \inst{2}  
\and T. Contini \inst{6}
\and S. Foucaud \inst{14}
\and P. Franzetti \inst{8}
\and L. Guzzo \inst{15}
\and O. Ilbert \inst{16}
\and A. Iovino \inst{15}
\and H.J. McCracken \inst{13,17}
\and C. Marinoni \inst{18}
\and A. Mazure \inst{9}
\and B. Meneux \inst{8,15}
\and R. Merighi   \inst{2} 
\and R. Pell\`o \inst{6}
\and A. Pollo \inst{9,19}
\and L. Pozzetti    \inst{2} 
\and M. Radovich \inst{20}
\and E. Zucca    \inst{2}
\and E. Hatziminaoglou \inst{21}
\and M. Polletta \inst{22}
\and M. Bondi \inst{10}
\and J. Brinchmann  \inst{23} 
\and O. Cucciati \inst{15,24}
\and S. de la Torre \inst{9}
\and L. Gregorini \inst{25}
\and Y. Mellier \inst{13,17}
\and P. Merluzzi \inst{20}
\and S. Temporin \inst{15}
\and D. Vergani \inst{8}
\and C.J. Walcher \inst{9}
}

\offprints{Angela Bongiorno, \email{angela.bongiorno@oabo.inaf.it}}

\institute{
Universit\`a di Bologna, Dipartimento di Astronomia - Via Ranzani 1,
I-40127, Bologna, Italy
\and 
INAF-Osservatorio Astronomico di Bologna - Via Ranzani 1, I-40127, Bologna, Italy
\and
Astrophysical Institute Potsdam, An der Sternwarte 16, D-14482
Potsdam, Germany
\and
Integral Science Data Centre, ch. d'\'Ecogia 16, CH-1290 Versoix
\and
Geneva Observatory, ch. des Maillettes 51, CH-1290 Sauverny, Switzerland
\and
Laboratoire d'Astrophysique de Toulouse/Tabres (UMR5572), CNRS, Universit\'e Paul Sabatier -
Toulouse III, Observatoire Midi-Pyri\'en\'ees, 14 av. E. Belin, F-31400 Toulouse, France 
\and 
Institute for Astronomy, University of Edinburgh, Royal Observatory, Edinburgh EH9 3HJ
\and
IASF-INAF - via Bassini 15, I-20133, Milano, Italy
\and
Laboratoire d'Astrophysique de Marseille, UMR 6110 CNRS-Universit\'e de
Provence,  BP8, 13376 Marseille Cedex 12, France
\and
IRA-INAF - Via Gobetti,101, I-40129, Bologna, Italy
\and
INAF-Osservatorio Astronomico di Roma - Via di Frascati 33,
I-00040, Monte Porzio Catone, Italy
\and
Max Planck Institut fur Astrophysik, 85741, Garching, Germany
\and
Institut d'Astrophysique de Paris, UMR 7095, 98 bis Bvd Arago, 75014
Paris, France
\and
School of Physics \& Astronomy, University of Nottingham, University Park, Nottingham, NG72RD, UK
\and
INAF-Osservatorio Astronomico di Brera - Via Brera 28, Milan, Italy
\and
Institute for Astronomy, 2680 Woodlawn Dr., University of Hawaii, 
Honolulu, Hawaii, 96822
\and	
Observatoire de Paris, LERMA, 61 Avenue de l'Observatoire, 75014 Paris, 
France
\and
Centre de Physique Th\'eorique, UMR 6207 CNRS-Universit\'e de Provence, 
F-13288 Marseille France
\and
Astronomical Observatory of the Jagiellonian University, ul Orla 171, 
30-244 Krak{\'o}w, Poland
\and
INAF-Osservatorio Astronomico di Capodimonte - Via Moiariello 16, I-80131, Napoli,
Italy
\and
Institute de Astrofisica de Canarias, C/ Via Lactea s/n, E-38200 La Laguna, Spain
\and
Center for Astrophysics \& Space Sciences, University of California, San Diego, La Jolla, CA 92093-0424, USA
\and
Centro de Astrofísica da Universidade do Porto, Rua das Estrelas,
4150-762 Porto, Portugal 
\and
Universit\'a di Milano-Bicocca, Dipartimento di Fisica - 
Piazza delle Scienze 3, I-20126 Milano, Italy
\and
Universit\`a di Bologna, Dipartimento di Fisica - Via Irnerio 46,
I-40126, Bologna, Italy
}

   \date{Received; accepted }

 
  \abstract{
In a previous paper (Gavignaud et al. 2006), we presented the type--1 Active
Galactic Nuclei (AGN) sample obtained from the first epoch data of the VIMOS-VLT
Deep Survey (VVDS). The sample consists of 130 faint, broad-line AGN with
redshift up to $z=5$ and $17.5<I_{AB}<24.0$, selected on the basis of their
spectra. 

In this paper we present the measurement of the Optical Luminosity Function up to 
$z=3.6$ derived from this sample, we compare our results with previous results
from brighter samples both at low and at high redshift.

Our data, more than one magnitude fainter than previous optical surveys, 
allow us to constrain the faint part of the luminosity function up to high
redshift. 

By combining our faint VVDS sample with the large
sample of bright AGN extracted from the SDSS DR3 (Richards et al., 2006b), 
we find that the model which better represents the combined luminosity functions, 
over a wide range of redshift and luminosity, is a luminosity dependent density 
evolution (LDDE) model, similar to those derived from the major X-surveys. Such a
parameterization allows the redshift of the AGN space density peak to change
as a function of luminosity and explains the excess of faint AGN that we find
at 1.0 $< z <$ 1.5. On the basis of this model we find, for the first time from
the analysis of optically selected samples, that the peak of the
AGN space density shifts significantly towards lower redshift going to lower
luminosity objects. This result, already found in a number of 
X-ray selected samples of AGN, is consistent with a scenario of ``AGN cosmic
downsizing'', in which the density of more luminous AGN, possibly associated to
more massive black holes, peaks earlier in the history of the Universe, 
than that of low luminosity ones.}


\titlerunning{The VVDS type--1 AGN sample: The faint end of the luminosity function}
\authorrunning{Bongiorno, A. et al.}

\maketitle
  
%
\section{Introduction}
Active Galactic Nuclei (AGN) are relatively rare objects that exhibit some of the most
extreme physical conditions and activity known in the universe. 

A useful way to statistically describe the AGN activity along the cosmic time
is through the study of their luminosity function, whose shape,
normalization and evolution can be used to derive constraints on models 
of cosmological evolution of black holes (BH). 
At z$\lesssim$2.5, the luminosity function of optically selected type--1 AGN has been
well studied since many years \citep{Boyle1988,Hewett1991,Pei1995,Boyle2000,
Croom2004}. It is usually described as a double power law, characterized by the 
evolutionary parameters $L^*(z)$ and $\Phi^*(z)$, which allow to distinguish 
between simple evolutionary models such as Pure Luminosity Evolution (PLE) and 
Pure Density Evolution (PDE). 
Although the PLE and PDE models should be mainly considered as mathematical 
descriptions of the evolution of the luminosity function, two different physical 
interpretations can be associated to them: either a small fraction of bright 
galaxies harbor AGN, and the luminosities of these sources change systematically with time 
(`luminosity evolution'), or all bright galaxies harbor AGN, but at any given time most 
of them are in `inactive' states.
In the latter case, the fraction of galaxies with AGN in an `active' state changes with 
time (`density evolution'). 
Up to now, the PLE model is the preferred
description for the evolution of optically selected QSOs, at least at low 
redshift ($z<2$).

Works on high redshift type--1 AGN samples \citep{Warren1994, Kennefick1995,Schmidt1995,
Fan2001,Wolf2003,Hunt2004} have shown that the number density of QSOs 
declines rapidly from $z \sim 3$ to $z \sim 5$. Since the size of complete and
well studied samples of QSOs at high redshift is still relatively small, the rate of this
decline and the shape of the high redshift luminosity function is  not yet as
well constrained as at low redshift. 
For example, \cite{Fan2001}, studying a sample of 39 luminous high redshift
QSOs at $3.6<z<5.0$, selected from the commissioning data of the Sloan 
Digital Sky Survey (SDSS), found that the slope of the bright end of the QSO luminosity function 
evolves with redshift, becoming flatter at high redshift, and that the QSO 
evolution from $z=2$ to $z=5$ cannot be described as a pure luminosity 
evolution. A similar result on the flattening at high redshift of the slope of
the luminosity function for luminous QSOs has been recently obtained by
\cite{Richards2006QLF} from the analysis of a much larger sample of SDSS QSOs
(but see \cite{Fontanot2007} for different conclusions drawn on the basis of
combined analysis of GOODS and SDSS QSOs).

At the same time, a growing number of observations at different redshifts, in
radio, optical and soft and hard X-ray bands, are suggesting  that also the
faint end slope evolves, becoming flatter at high redshift \citep{Page1997,
Miyaji2000,Miyaji2001,LaFranca2002,Cowie2003,Ueda2003,Fiore2003,Hunt2004,
Cirasuolo2005,Hasinger2005}. This evolution, now dubbed as  ``AGN cosmic
downsizing'' is described either as a direct evolution in the faint end slope or
as ``luminosity dependent density evolution" (LDDE), and it has been the subject
of many speculations since it implies that the space density of low luminosity
AGNs peaks at lower redshift than that of bright ones.

It has been observed that, in addition to the well known local scale relations 
between the black hole (BH) masses and the properties of their host galaxies
\citep{Kormendy1995,Magorrian1998,Ferrarese2000}, also the galaxy spheroid 
population follows a similar pattern of ``cosmic
downsizing'' \citep{Cimatti2006}. Various models have been proposed to explain 
this common evolutionary trend in AGN and spheroid galaxies. The majority of
them propose that the feedback from the black hole growth plays a key role in
determining the BH-host galaxy relations \citep{Silk1998,DiMatteo2005} and the
co-evolution of black holes and their host galaxies. Indeed, AGN feedback can
shut down the growth of the most massive systems steepening the bright end slope
\citep{Scannapieco2004}, while the feedback-driven QSO decay determines the
shape of the faint end of the QSO LF \citep{Hopkins2006b}.

This evolutionary trend has not been clearly seen yet with optically selected
type--1 AGN samples. By combining results from low and high redshifts, it is clear from
the studies of optically selected samples that the cosmic QSO evolution shows a
strong increase of the activity from $z\sim 0$ out to $z\sim 2$, reaches a
maximum around $z \simeq 2-3$ and then declines, but the shape of the turnover 
and the redshift evolution of the peak in activity as a function of luminosity
is still unclear.

Most of the optically selected type--1 AGN samples studied so far are obtained through 
various color selections of candidates, followed by
spectroscopic confirmation (e.g. 2dF, \citealt{Croom2004} and SDSS, 
\citealt{Richards2002}), or grism and slitless spectroscopic surveys. These
samples are expected to be highly complete, at least for luminous type--1 AGN, at either
 $z \leq 2.2$ or $z \geq 3.6$, where type--1 AGN show conspicuous colors in broad band
color searches, but less complete in the redshift range $2.2 \leq z \leq 3.6$
(Richards et al. 2002).  

An improvement in the multi-color selection in optical bands is through
the simultaneous use of many broad and medium band filters as in
the COMBO-17 survey \citep{Wolf2003}. This survey is the only optical survey
so far which, in addition to covering a redshift range large enough to see the
peak of AGN activity, is also deep enough to sample up to high redshift type--1 AGN with
luminosity below the break in the luminosity function. However, only photometric
redshifts are available for this sample and, because of their selection
criteria, it is incomplete for objects with a small ratio between the nuclear
flux and the total host galaxy flux and for AGN with anomalous colors, such as,
for example, the broad absorption line (BAL) QSOs , which have on average redder
colors and account for $\sim$ 10 - 15 \% of the overall AGN population \citep{Hewett2003}.

The VIMOS-VLT Deep Survey \citep{LeFevre2005} is a spectroscopic survey in which the target
selection is purely flux limited (in the I-band), with no additional selection
criterion. This allows the selection of a spectroscopic type--1 AGN sample free of color and/or 
morphological biases
in the redshift range z $>$ 1. An obvious advantage of such a selection is
the possibility to test the completeness of the most current surveys 
\citep[see][ Paper I]{Gavignaud2006}, based on morphological and/or color pre-selection, 
and to study the evolution of type--1 AGN activity in a large redshift range.

In this paper we use the type-1 AGN sample selected from the VVDS to derive the luminosity 
function in the redshift range $1<z<3.6$. The VVDS type--1 AGN sample is more than one magnitude deeper 
than any previous optically selected sample and allow thus to explore the faint part 
of the luminosity function. Moreover, by combining this LF with measurement of the LF in 
much larger, but very shallow, surveys, we find an analytical form to dercribe, in a large luminosity 
range, the evolution of type-1 AGN in the redshift range 0$< z <$4.    
The paper is organized as follows: in Section 2 and 3 we describe the sample and
its color properties. In Section 4 we present the method used to derive the
luminosity function, while in Section 5 we compare it with previous works both
at low and high redshifts. The bolometric LF and the comparison with the results
derived from samples selected in different bands (from X-ray to IR) is then 
presented in Section 6. The derived LF  fitting models are presented in Section 7 
while the AGN activity as a function of redshift is shown in Section 8.  
Finally in section 9 we summarize our results.
Throughout this paper, unless stated otherwise, we assume a cosmology with 
$\Omega_{m}$ = 0.3, $\Omega_{\Lambda}$ = 0.7 and H$_{0}$ = 70 km s$^{-1}$
Mpc$^{-1}$.

\section{The sample}
Our AGN sample is extracted from the first epoch data of the VIMOS-VLT Deep Survey,
performed in 2002 \citep{LeFevre2005}.

\begin{figure}                                                      
\begin{center}                                                      
\includegraphics[height=7.0cm,width=7.0cm]{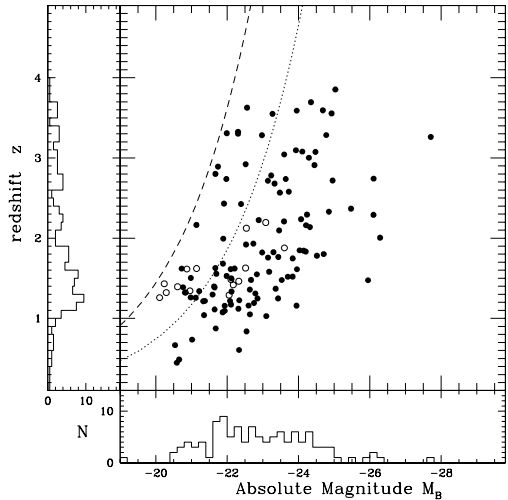} 
\caption{Distribution of absolute magnitudes and redshifts of the total AGN
sample. Open circles are the objects with ambiguous redshift, shown at all their
possible z values. The dotted and dashed lines represent the magnitude limits of
the samples: $I_{AB} < 22.5$ for the wide sample and  $I_{AB} < 24.0$ for the deep sample.}
\label{fig:MBz}                                                     
\end{center}                                 
\end{figure}
 
The VVDS is a spectroscopic survey designed to measure
about 150,000 redshifts of galaxies, in the redshift range $0<z<5$, selected,
nearly randomly, from an imaging survey (which consists of observations in U, B,
V, R and I bands and, in a small area, also K-band) designed for this purpose.
Full details about VIMOS photometry can be found in \cite{LeFevre2004a},
\cite{McCracken2003}, \cite{Radovich2004} for the U-band and \cite{Iovino2005}
for the K-band. In this work we will as well use the Galex UV-catalog 
\citep{Arnouts2005,Schiminovich2005}, the $u^*$,$g'$,$r'$,$i'$,$z'$ photometry
obtained in the frame of the Canada-France-Hawaii Legacy Survey
(CFHTLS)\footnote{www.cfht.hawaii.edu/Science/CFHLS}, UKIDSS 
\citep{Lawrence2006}, and the Spitzer Wide-area InfraRed Extragalactic survey
(SWIRE) \citep{Lonsdale2003,Lonsdale2004}. The spectroscopic VVDS survey consists of a deep
and a wide survey and it is based on a simple selection function. The sample is
selected only on the basis of the I band magnitude: $17.5<I_{AB}<22.5$ for the
wide and $17.5<I_{AB}<24.0$ for the deep sample. For a detailed description of
the spectroscopic survey strategy and the first epoch data see \cite{LeFevre2005}.

Our sample consists of 130 AGN with $0<z<5$, selected in 3 VVDS fields
(0226-04, 1003+01 and 2217-00) and in the Chandra Deep Field South 
\citep[CDFS,][]{LeFevre2004cdfs}. 
All of them are selected as AGN
only on the basis of their spectra, irrespective of their morphological or
color properties. In particular, we selected them on the basis of the presence
of at least one broad emission line. We discovered 74 of them in the deep fields
(62 in the 02h field and 12 in the CDFS) and 56 in the wide fields (18 in the 
10h field and 38 in the 22h field). This represents an unprecedented complete 
sample of faint AGN, free of morphological or color selection bias. 
The spectroscopic area covered by the First Epoch Data is 0.62 $deg^2$ in the
deep fields (02h field and CDFS) and 1.1 $deg^2$ in the wide fields (10h and 
22h fields). 

To each object we have assigned a value for the spectroscopic
redshift and a spectroscopic quality flag which quantifies our confidence level
in that given redshift. As of today, we have 115 AGN with secure redshift,
and 15 AGN with two or more possible values for the redshift. For these objects,
we have two or more possible redshifts because only one broad emission line,
with no other narrow lines and/or additional features, is detected in the
spectral wavelength range adopted in the VVDS (5500 - 9500 \AA) 
(see Figure 1 in Paper I). For all of them, however, a {\it best solution} is proposed. 
In the original VVDS AGN sample, the number of AGN with this redshift degeneracy was 42. 
To solve this problem, we have first looked for the objects already
observed in other spectroscopic surveys in the same areas, solving the redshift for 3 of them.
For the remaining objetcs, we performed a spectroscopic
follow-up with FORS1 on the VLT Unit Telescope 2 (UT2). With these additional 
observations we found a
secure redshift for 24 of our AGN with ambiguous redshift
determination  and, moreover, we found that our proposed best solution was the correct one
in $\sim$ 80\% of the cases. On the basis of this result, we decided to use, in
the following analysis, our best estimate of the redshift for the small remaining
fraction of AGN with ambiguous redshift determination (15 AGN).

\begin{figure}                                                     
\begin{center}                                                     
\includegraphics[height=6.0cm,width=8.0cm]{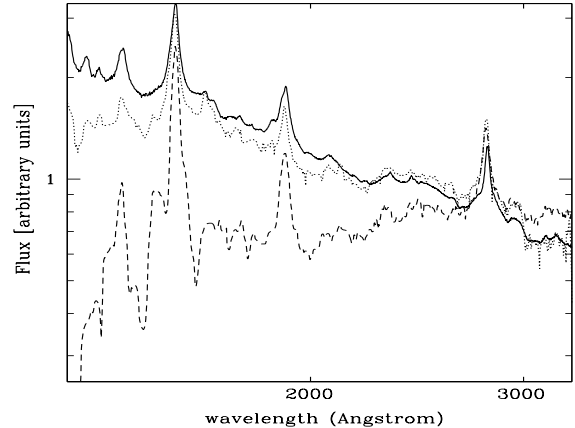} 
\caption{Composite spectra derived for our AGN with secure redshift in the 02h
field, divided in a ``bright'' (19 objects at M$_{1450}<$-22.15, dotted curve) 
and a ``faint'' (31 objects at M$_{1450}>$-22.15, dashed curve) sample. We consider
here only AGN with $z>1$ (i.e. the AGN used in to compute the luminosity
function). The SDSS composite spectrum is shown with a solid line for comparison. }
\label{fig:comp_b_f}                                               
\end{center}                                
\end{figure}

In Figure \ref{fig:MBz} we show the absolute B-magnitude and the redshift 
distributions of the sample. As shown in this Figure, our sample spans a 
large range of luminosities and consists of both Seyfert galaxies (M$_B >$-23; $\sim$59\%) 
and QSOs (M$_B <$-23; $\sim$41\%). 
A more detailed and exhaustive description of the properties of the AGN sample 
is given in Paper I \citep{Gavignaud2006} and the complete list of
BLAGN in our wide and deep samples is available as an electronic Table 
in Appendix of \cite{Gavignaud2006}.

\section{Colors of BLAGNs} \label{sez:colors}

As already discussed in Paper I, the VVDS AGN sample shows, on average, redder
colors than those expected by comparing them, for example, with the color track
derived from the SDSS composite spectrum \citep{VandenBerk2001}. In Paper I we proposed three possible
explanations: (a) the contamination of the host galaxy is reddening the observed
colors of faint AGN; (b) BLAGN are intrinsically redder when they are faint; (c)
the reddest colors are due to dust extinction. On the basis of the
statistical properties of the sample, we concluded that hypothesis (a) was
likely to be the more correct, as expected from the faint absolute magnitudes
sampled by our survey, even if hypotheses (b) and (c) could not be ruled out.

In Figure \ref{fig:comp_b_f} we show the composite spectra derived from the
sample of AGN with secure redshift in the 02h field, divided in a ``bright''
and a ``faint'' sample at the
absolute magnitude $M_{1450}=-22.15$. We consider here only AGN with $z>1$, which
correspond to the AGN used in Section \ref{sez:LF} to compute the luminosity
function.  The choice of the reference wavelength for the absolute magnitude,
$\lambda= 1450$ \AA , is motivated by our photometric coverage. In fact, for most
of the objects it is possible to interpolate $M_{1450}$ directly from the
observed magnitudes. In the same plot we show also the SDSS composite spectrum
(solid curve) for comparison. Even if also the "bright" VVDS composite (dotted
curve) is somewhat redder than the SDSS one, it is clear from this plot that the
main differences occur for faintest objects (dashed curve).
 
\begin{figure}
\begin{center}                                                      
\includegraphics[height=11.5cm,width=8.0cm]{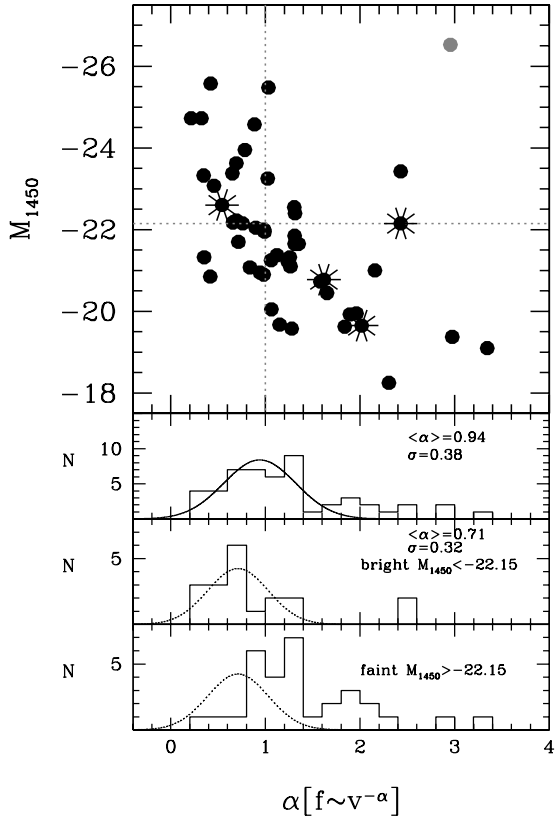} 
\caption{{\it Upper Panel}: Distribution of the spectral index $\alpha$ as a function of M$_{1450}$
for the same sample of AGN as in Figure \ref{fig:comp_b_f}.
The spectral index is derived here by fitting a simple power law
$f(\nu)=\nu^{-\alpha}$  to our photometric data points. 
Asterisks are AGN morphologically classified as extended and the grey point is a BAL AGN.
{\it Bottom Panels:} Distribution of the spectral index $\alpha$ for the same sample of AGN. 
All the AGN in this sample are shown in the first of the three panels, while the AGN in 
the ``bright'' and ``faint'' sub--samples are shown in the second and third panel, respectively. 
The dotted curve in the second panel corresponds to the gaussian fit of the
bright sub--sample and it is reported also in the third panel to highlight the
differences in the $\alpha$ distributions of the two sub-samples.}
\label{fig:alpha_M1450} 
\end{center}                                 
\end{figure}

A similar result is shown for the same sample in the upper panel of Figure 
\ref{fig:alpha_M1450}, where we plot the spectral index $\alpha$ as a function 
of the AGN luminosity. The spectral index is derived here by fitting a simple 
power law $f(\nu)=\nu^{-\alpha}$  to our
photometric data points. This analysis has been performed only on the 02h deep
sample, since for the wide sample we do not have enough photometric coverage to
reliably derive the spectral index. Most of the AGN with $\alpha > 1$ are fainter
than $M_{1450}=-22.15$, showing that, indeed, the faintest objects have on
average redder colors than the brightest ones. The outlier (the brightest object
with large $\alpha$, i.e. very red colors, in the upper right corner of the plot)
is a BAL AGN.

The three bottom panels of Figure \ref{fig:alpha_M1450} show the histograms of the resulting power law
slopes for the same AGN sample. The total sample is plotted in the first panel,
while the bright and the faint sub-samples are plotted in the second and third
panels, respectively. A Gaussian curve with $<\alpha>=0.94$ and dispersion $\sigma
= 0.38$ is a good representation for the distribution of about 80\% (40/50) of the
objects in the first panel. In addition, there is a significant tail ($\sim$ 20\%)
of redder AGN with slopes in the range from 1.8 up to $\sim$ 3.0. The average slope
of the total sample ($\sim$ 0.94) is redder than the fit to the SDSS composite
($\sim$ 0.44). Moreover, the distribution of $\alpha$ is shifted toward much larger
values (redder continua) than the similar distribution in the SDSS sample
\citep{Richards2003}. For example, only 6\% of the objects in the SDSS sample
have $\alpha>1.0$, while this percentage is 57\% in our sample.

The differences with respect to the SDSS sample can be partly due to the differences
in absolute magnitude of the two samples ($M_i <$-22.0 for the SDSS sample \citep{Schneider2003} 
and M$_{B} <$-20.0 for the VVDS sample). 
In fact,  if we consider the VVDS ``bright'' sub-sample, the average spectral index $<\alpha>$ 
becomes $\sim$ 0.71, which is closer to the SDSS value (even if it is still somewhat redder), 
and only two objects ($\sim$8\% of the sample) show values not consistent with a gaussian
distribution with $\sigma \sim$0.32. Moreover, only 30\% of this sample have
$\alpha>1.0$.

Most of the bright SDSS AGNs with $\alpha>1$ are interpreted by \cite{Richards2003} to be
dust-reddened, although a fraction of them is likely to be due to intrinsically
red AGN \citep{Hall2006}. At fainter magnitude one would expect both a larger
fraction of dust-reddened objects (in analogy with indications from the X-ray data
\citep{Brandt2000,Mushotzky2000} and a more significant contamination from the host
galaxy.

We have tested these possibilities by examining the global Spectral Energy
Distribution (SED) of each object and fitting the observed fluxes $f_{obs}$ with a
combination of AGN and galaxy
emission, allowing also for the possibility of extinction of the AGN flux. 
Thanks to the multi-wavelength coverage in the deep field in which we have, in
addition to VVDS bands, also data from GALEX, CFHTLS, UKIDSS and SWIRE, we can
study the spectral energy distribution of the single objects. In particular, we 
assume that:
\begin{equation}
f_{obs}=c_1f_{AGN} \cdot 10^{-0.4 \cdot A_{\lambda}}+c_2f_{GAL}
\end{equation}
and, using a library of galaxy and AGN templates, we find the best parameters $c_1$,
$c_2$ and $E_{B-V}$ for each object.
We used the AGN SED derived by \cite{Richards2006SED}
with an SMC-like dust-reddening law \citep{Prevot1984} with the form 
$A_{\lambda}/E_{B-V} =1.39 \lambda_{\mu m}^{-1.2}$, and a library of galaxies
template by \cite{Bruzual2003}.

\begin{figure}
\begin{center}                                                      
\includegraphics[height=10.0cm,width=9.0cm]{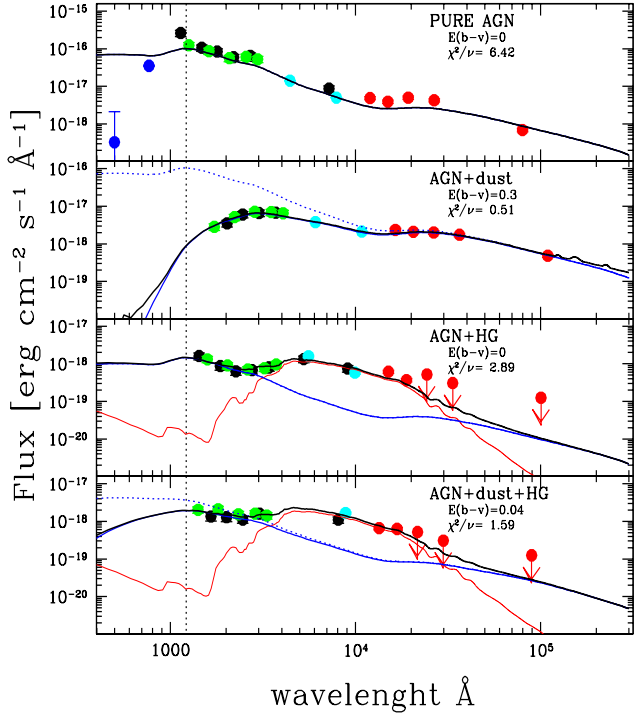} 
\caption{Four examples of different decompositions of the observed SEDs of our
objects. Since for $\lambda <$ 1216 \AA, corresponding to the 
Ly$\alpha$ line, the observed flux is expected to decrease because of intervening
absorption, all the photometric data at $\lambda<$1216 \AA\ are not considered in the fitting. The only requested constraint is that they 
lie below the fit. The four fits shown in this Figure correspond, from top to
bottom, to pure-AGN, dust-extincted AGN, AGN and host galaxy, dust-extincted AGN
and host galaxy. The dotted line corresponds to the AGN template before applying the 
extinction law, while the solid blue line corresponds to the same template, but extincted 
for the given $E_{B-V}$. The red line (third and fourth panel) corresponds to the galaxy template 
and, finally, the black line is the resulting best fit to the SED. 
Arrows correspond to 5$\sigma$ upper limits in case 
of non detection in the IR.}
\label{fig:AGNextHG} 
\end{center}                                 
\end{figure}

We found that for $\sim$37\% of the objects, the observed flux is fitted by a typical 
AGN power law (pure AGN), while 44\% of the sources require the presence of a contribution 
from the host galaxy to reproduce the observed flux. Only 4\% of the objects are fitted by pure 
AGN + dust, while the remaining 15\% of objects require
instead both contributions (host galaxy contamination and presence of dust).
As expected, if we restrict the analysis to the bright sample, the percentage of 
pure AGN increases to 68\%, with the rest of the objects requiring either some contribution 
from the host galaxy ($\sim$21\%) or the presence of dust oscuration ($\sim$11\%).\\    
In Figure \ref{fig:AGNextHG} we show 4 examples of the resulting fits:
(i) pure AGN; (ii) dust-extincted AGN; (iii) AGN contaminated by the host galaxy;
(iv) dust-extincted AGN and contaminated by the host galaxy.
The dotted line corresponds to the AGN template before applying the 
extinction law, while the solid blue line corresponds to the same template, 
but extincted for the given $E_{B-V}$; the red 
line corresponds to the galaxy template and, finally, the black line is the resulting 
best fit to the SED.
The host galaxy contaminations will be taken into account in the computation of the AGN 
absolute magnitude for the luminosity function.

\section{Luminosity function} \label{sez:LF}

\subsection{Definition of the redshift range} \label{sez:redrange}

For the study of the LF we decided to exclude AGN with $ z \leq 1.0$. This choice is due 
to the fact that for $ 0.5 \leq z \leq 1.0$ the only visible broad line in 
the VVDS spectra is H$\beta$ (see Figure 1 of Paper I). This means that all objects
with narrow or almost narrow H$\beta$ and broad H$\alpha$ (type 1.8, 1.9 AGN; see
\citealt{Osterbrock1981}) would not be included in our sample, because we include
in the AGN sample all the objects with at least one visible broad line. Since at
low luminosities the number of intermediate type AGN is not negligible, this
redshift bin is likely to be under-populated and the results would not be
meaningful.

At $z < 0.5$, in principle we have less problems, because also H$\alpha$ is within
the wavelength range of the VVDS spectra, but, since at this low redshift, 
our sampled volume is relatively small and QSOs rare, only 3 objects have secure redshifts
in this redshift bin in the current sample. For these reasons, our luminosity
function has been computed only for $z > 1.0$ AGN. As already mentioned in Section
2, the small fraction of objects with an ambiguous redshift determination have been
included in the computation of the luminosity function assuming that our best
estimate of their redshift is correct.\\
The resulting sample used in the computation of the LF consists thus of 121 objects 
at 1$< z <$4.

\subsection{Incompleteness function} \label{sez:incompletenessfunction}

Our incompleteness function is made up of two terms linked, respectively, to the
selection algorithm and to the spectral analysis: the Target Sampling Rate (TSR) and the 
Spectroscopic Success Rate (SSR) defined following \citet{Ilbert2005}.

{\it The Target Sampling Rate}, namely the ratio between the observed sources and the
total number of objects in the photometric catalog, quantifies the incompleteness
due to the adopted spectroscopic selection criterion. 
The TSR is similar in the wide and deep sample and runs from 20\% to 30\%.

{\it The Spectroscopic Success Rate} is the probability of a
spectroscopically targeted object to be securely identified.
It is a complex function of the BLAGN redshift, apparent magnitude and
intrinsic spectral energy distribution and it has been estimated by 
simulating 20 Vimos pointings, for a total of 2745 spectra.

Full details on TSR and SSR can be found in Paper I \citep{Gavignaud2006}. 
We account for them by computing for
each object the associated weights $w^{tsr}=1/TSR$ and  $w^{ssr}=1/SSR$; the total 
weighted contribution of each object to the luminosity function is then
the product of the derived weights ($w^{tsr} \times w^{ssr}$).

\subsection{Estimate of the absolute magnitude} \label{sez:AbsoluteMagnitude}

We derived the absolute magnitude in the reference band from the apparent magnitude 
in the observed band as:
\begin{equation}
M = m_{obs} -5log_{10}(dl(z))-25-k
\end{equation}
where M is computed in the band in which we want to compute the luminosity function, 
$m_{obs}$ is the observed band from which we want to calculate it, dl(z) is the luminosity 
distance expressed in Mpc and k is the k-correction in the reference band.
To make easier the comparison with previous results in the literature, we computed 
the luminosity function in the B-band.

To minimize the uncertainties in the adopted k-correction, $m_{obs}$ for each object
should be chosen in the observed band which is sampling the rest-wavelength closer
to the band in which the luminosity function is computed. For our sample, which
consists only of $z>1$ objects, the best bands to use to compute the B-band 
absolute magnitudes should be respectively the I-, J- and K-bands going to higher redshift.
Since however, the only observed band available for the entire sample (deep and wide), is 
the I-band, we decided to use it for all objects to compute the B-band magnitudes. 
This means that for $z \gsim$ 2, we introduce an uncertainty in the absolute magnitudes 
due to the k-correction.
We computed the absolute magnitude considering the template derived from the 
SDSS sample \citep{VandenBerk2001}.

\begin{figure}
\begin{center}                                                     
\includegraphics[height=7cm,width=7cm]{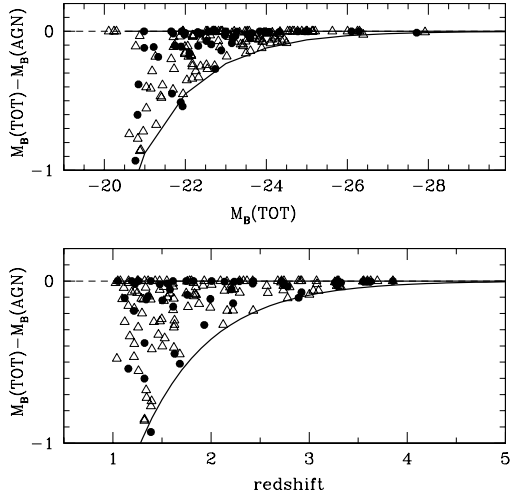} 
\caption{Real (full circles; AGN in the deep sample) and simulated (open triangles;
AGN in the wide sample) B-band absolute
magnitude differences as a function of M$_B$(TOT) (upper panel) and redshift
(bottom panel). M$_B$(TOT) is the absolute magnitude 
computed considering the total observed flux, while M$_B$(AGN) is the absolute
magnitude computed after subtracting the host-galaxy contribution.}
\label{fig:DeltaM}
\end{center}                                
\end{figure}

As discussed in Section \ref{sez:colors}, the VVDS AGN sample shows redder colors
than those typical of normal, more luminous AGN and this can be due to the
combination of the host galaxy contribution and the presence of dust. Since, in this
redshift range, the fractional contribution from the host galaxies is expected 
to be more significant in the I-band than in bluer bands, the luminosity derived
using the I-band observed  magnitude could, in some cases, be somewhat overestimated
due to the contribution of the host galaxy component. 

We estimated the possible impact of this effect on our results in the following way.
From the results of the analysis of the SED of the single objects in the deep sample 
(see Section \ref{sez:colors}) we computed for each object the difference 
$m_{I}(TOT)-m_{I}(AGN)$ and, consequently, $M_{B}(TOT)-M_{B}(AGN)$. 
This could allow us to derive the LF using directly the derived $M_{B}(AGN)$, 
resolving the possible bias introduced by the host galaxy contamination.
These differences are shown as full circles in Figure \ref{fig:DeltaM} as a function of
absolute magnitude (upper panel) and redshift (lower panel). For most of the objects
the resulting differences between the total and the AGN magnitudes are small 
($\Delta$M$\leq$0.2). However, for a not negligible fraction of the faintest 
objects (M$_B \geq$-22.5, $z \leq$2.0) 
these differences can be significant (up to $\sim$1 mag).
For the wide sample, for which the more restricted photometric coverage does not
allow a detailed SED analysis and decomposition, we used simulated differences to derive the M$_B$(AGN). 
These simulated differences have been derived through a Monte Carlo 
simulation on the basis of the bivariate distribution $\Delta$M(M,z) estimated from
the objects in the deep sample. 
$\Delta$M(M,z) takes into account the probability distribution of $\Delta$M as a function of M$_B$ 
and z, between 0 and the solid line in Figure \ref{fig:DeltaM} derived as the envelope suggested 
by the black dots. 
The resulting simulated differences for the objects
in the wide sample are shown as open triangles in the two panels of Figure \ref{fig:DeltaM}.

The AGN magnitudes and the limiting magnitudes of the samples have been corrected
also for galactic extinction on the basis of the mean extinction values $E(B-V)$ in
each field derived from \cite{Schlegel1998}. Only for the 22h field, where the
extinction is highly variable across the field, we used the extinction on 
the basis of the position of individual objects. The resulting corrections in the
I-band magnitude are $A_{I} \simeq 0.027$ in the 2h and 10h fields and 
$A_{I} = 0.0089$ in the CDFS field, while the average value in the 22h field is 
$A_{I} = 0.065$. These corrections have been applied also to the limiting magnitude
of each field.

\begin{figure*}
\begin{center}                               
\includegraphics[height=9cm,width=0.9\linewidth]{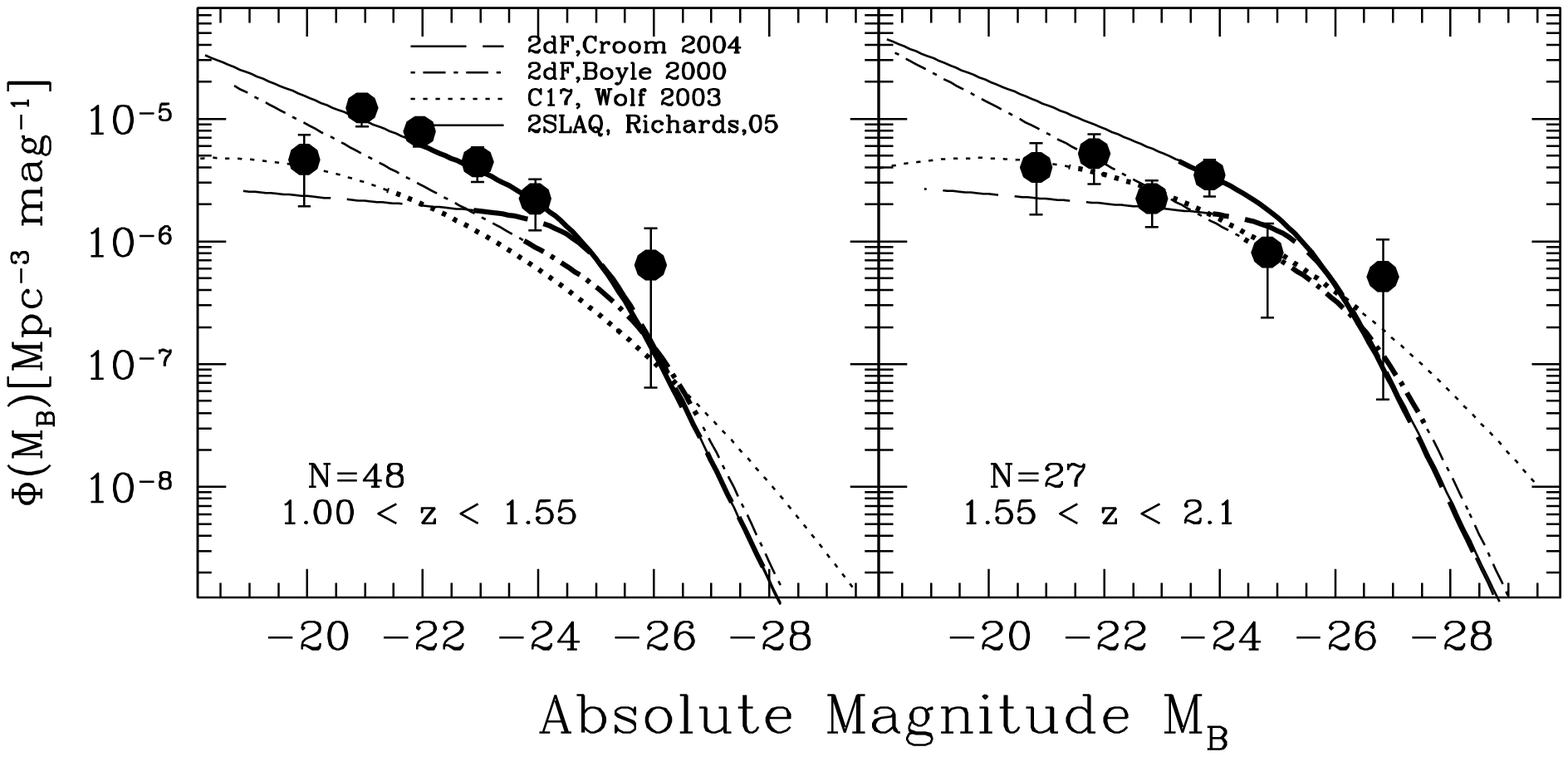} 
\caption{Our rest-frame B-band luminosity function, derived in the redshift bins
$1.0<z<1.55$ and $1.55<z<2.1$, compared with the 2dFQRS 
\citep{Croom2004,Boyle2000}, COMBO-17 data \citep{Wolf2003} and with the 2dF-SDSS (2SLAQ) data 
\citep{Richards2005}. The curves in the Figure show the PLE fit models derived by these authors.
The thick parts of the curves correspond to the luminosity range covered by the
data in each sample, while the thin parts are extrapolations based on the best fit parameters of the
models.}
\label{fig:low}    
\end{center}
\end{figure*}

\subsection{The 1/V$_{max}$ estimator}

We derived the binned representation of the luminosity function using the usual
$1/V_{max}$ estimator \citep{Schmidt1968}, which gives the space density
contribution of individual objects. The luminosity function, for each redshift
bin ($z-\Delta z/2$ ; $z+\Delta z/2$), is then computed as:
\begin{equation}
\Phi(M) = \frac{1}{\Delta M} \sum_{M-\Delta M/2}^{M+\Delta M/2}\frac{w^{tsr}_i 
w^{ssr}_i}{V_{max,i}}
\end{equation}
where $V_{max,i}$ is the comoving volume within which the $i^{th}$ object would
still be included in the sample. $w_i^{tsr}$ and $w_i^{ssr}$ are respectively
the inverse of the TSR and of the SSR, associated to the $i^{th}$ object. 
The statistical uncertainty on $\Phi$(M) is given by  \cite{Marshall1983a}:

\begin{equation}
\sigma_{\phi} = \frac{1}{\Delta M} \sqrt {\sum_{M-\Delta M/2}^{M+\Delta M/2}
\frac {(  w_i^{tsr}  w^{ssr}_i)^2}{V_{max,i}^{2}}} 
\end{equation}

We combined our samples at different depths using the method proposed by 
\cite{Avni1980}. In this method it is assumed that each object, characterized by 
an observed redshift z$_i$ and intrinsic
luminosity L$_i$, could have been found in any of the survey areas for which its
observed magnitude is brighter than the corresponding flux limit.
This means that, for our total sample, we consider an area of:
$$
\Omega_{tot}(m) = \Omega_{deep} + \Omega_{wide} = \mbox{1.72 deg$^2$}  
\qquad \mbox{for} \quad  17.5 < I_{AB} < 22.5
$$
and
$$
\Omega_{tot}(m) = \Omega_{deep} = \mbox{0.62 deg$^2$}          
\qquad  \mbox{   for  } \quad  22.5 < I_{AB} < 24.0
$$

The resulting luminosity functions in different redshift ranges are plotted in
Figure \ref{fig:low} and \ref{fig:highz}, where all bins which contain at least
one object are plotted. 
The LF values, together with their 
1$\sigma$\ errors and the numbers of objects in each absolute magnitude bin are
presented in Table \ref{tab:lf}.
The values reported in Table \ref{tab:lf} and plotted in Figures \ref{fig:low} and 
\ref{fig:highz}  are not corrected for the host galaxy contribution.
We have in fact a posteriori verified that, even if the differences between the 
total absolute magnitudes and the magnitudes corrected for the host galaxy 
contribution (see Section \ref{sez:AbsoluteMagnitude}) can be
significant for a fraction of the faintest objects, the resulting luminosity functions 
computed by using these two sets of absolute magnitudes are not significantly different. 
For this reason and for a more direct comparison with previous works, the results on the luminosity
function presented in the next section are those obtained using the total
magnitudes.

\section{Comparison with the results from other optical surveys}  

We derived the luminosity function in the redshift range 1.0$< z <$3.6 and we compared it with the
results from other surveys at both low and high redshift.

\subsection{The low redshift luminosity function}

In Figure \ref{fig:low} we present our luminosity function up to $z=2.1$. 
The Figure show our LF data points (full circles) derived in
two redshift bins: $1.0<z<1.55$ and $1.55<z<2.1$ compared with the LF fits derived 
from the 2dF QSO sample by \cite{Croom2004} and by \cite{Boyle2000}, with the COMBO-17 sample
by \cite{Wolf2003}, and  with
the 2dF-SDSS (2SLAQ) LF fit by \cite{Richards2005}. In each panel the curves,
computed for the average z of the redshift range, correspond to a double power
law luminosity function in which the evolution with redshift is characterized
by a pure luminosity evolution modeled as $M^*_b(z)=M^*_b(0)-2.5(k_1z+k_2z^2)$. 
Moreover, the thick parts of the curves show the luminosity range covered by 
the data in each of the comparison samples, while the thin parts are
extrapolation based on the the best fit parameters of the models.

We start considering the comparison with the 2dF and the COMBO-17 LF fits. As
shown in Figure \ref{fig:low}, our bright LF data points connect rather smoothly
to the faint part of the 2dF data. However, our sample is more than two
magnitudes deeper than the 2dF sample. For this reason, a comparison at low
luminosity is possible only with the extrapolations of the LF fit.
At $z>1.55$, while the Boyle's model fits well our faint LF data points, the
Croom's extrapolation, being very flat, tends to underestimate our low
luminosity data points. At $z<1.55$ the comparison is worse: as in the higher
redshift bin, the Boyle's model fits our data better than the Croom's one
but, in this redshift bin, our data points show an excess at low luminosity
also with respect to Boyle's fit. This trend is similar to what shown also by
the comparison with the fit of the COMBO-17 data which, differently from the
2dF data, have a low luminosity limit closer to ours: at $z>1.55$ the agreement 
is very good, but in the first redshift bin our data show again an excess at
low luminosity. This excess is likely due to the fact that, because of its
selection criteria, the COMBO-17 sample is expected to be significantly
incomplete for objects in which the ratio between the nuclear flux and the
total host galaxy flux is small.
Finally, we compare our data with the 2SLAQ fits derived by
\cite{Richards2005}. The 2SLAQ data are derived from a sample of AGN selected
from the SDSS, at $18.0 <g< 21.85$ and $z<3$, and observed with the 2-degree
field instrument. Similarly to the 2dF sample, also for this sample the LF is
derived only for $z<2.1$ and $M_B<-22.5$. The plotted dot-dashed curve
corresponds to a PLE model in which they fixed most of the parameters of the
model at the values found by \cite{Croom2004}, leaving to vary only the faint
end slope and the normalization constant $\Phi^*$. In this case, the agreement 
with our data points at $z<1.55$ is very good also at low luminosity. 
The faint end slope found in this case is $\beta=-1.45$, which is similar to 
that found by \cite{Boyle2000} ($\beta=-1.58$) and significantly steeper than that
found by \cite{Croom2004} ($\beta=-1.09$). At $z>1.55$, the \cite{Richards2005}
LF fit tends to overestimate our data points at the faint end of the LF, which
suggest a flatter slope in this redshift bin.

The first conclusion from this comparison is that, at low redshift (i.e. $z<2.1$),
the data from our sample, which is $\sim$2 mag fainter than the previous 
spectroscopically confirmed samples, are not well fitted simultaneously in the
two analyzed redshift bins by the PLE models derived from the previous samples.
Qualitatively, the main reason for this appears to be the fact that our
data suggest a change in the faint end slope of the LF, which appears to flatten
with increasing redshift. This trend, already highlighted by previous X-ray
surveys \citep{LaFranca2002,Ueda2003,Fiore2003} 
suggests that a simple PLE parameterization may not be a good representation of
the evolution of the AGN luminosity function over a wide range of redshift and
luminosity. Different model fits will be discussed in Section 7. 
                                  
\begin{figure}
\begin{center}                               
 \includegraphics[height=9cm,width=9cm]{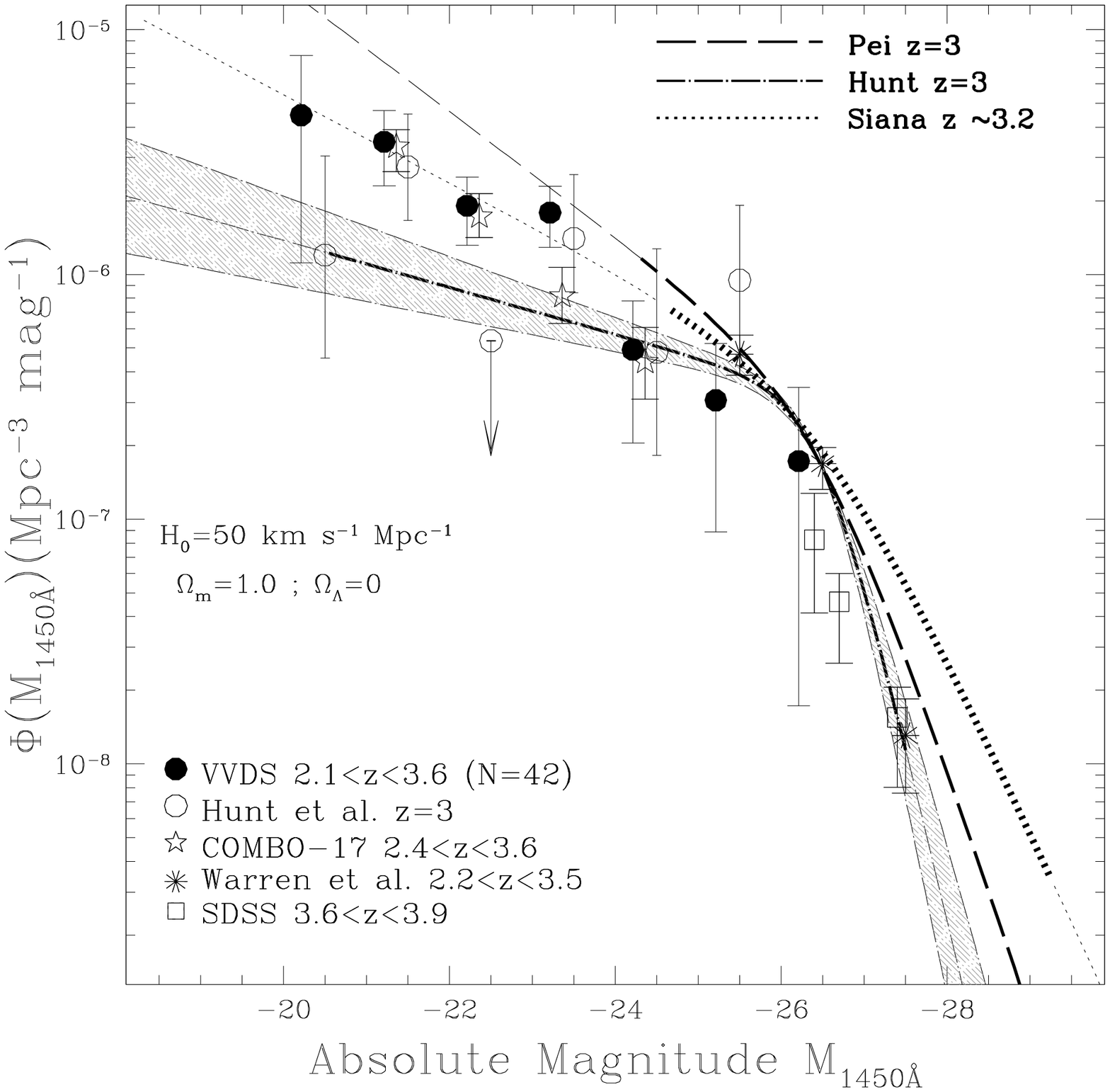}
\caption{Our luminosity function, at 1450 \AA\ rest-frame, in the redshift
range 2.1$<$z$<$3.6,  compared with data
from other high-z samples (\cite{Hunt2004} at $z=3$; Combo-17 data from 
\cite{Wolf2003} at $2.4<z<3.6$; data from \cite{Warren1994} at $2.2<z<3.5$
and the SDSS data from \cite{Fan2001}). The SDSS data points at 3.6$<z<$3.9
have been evolved to z=3 using the luminosity evolution of \cite{Pei1995} as in
 \cite{Hunt2004}. The curves show some model fits in which the thick parts
of the curves correspond to the luminosity range covered by the data samples, 
while the thin parts are model extrapolation. For this plot, an $\Omega_m=1$,
$\Omega_{\Lambda}=0$, $h=0.5$ cosmology has been assumed for comparison with
the previous works.}
\label{fig:highz}    
\end{center}                                 
\end{figure}

\subsection{The high redshift luminosity function}

The comparison of our LF data points for 2.1$< z <$3.6 (full circles) with the
results from other samples in similar redshift ranges is shown in Figure 
\ref{fig:highz}. In this Figure an $\Omega_m=1$, $\Omega_{\Lambda}=0$, $h=0.5$
cosmology has been assumed for comparison with previous works, and the absolute
magnitude has been computed at 1450 \AA. As before, the thick parts of the
curves show the luminosity ranges covered by the various data samples, while
the thin parts are model extrapolations. In terms of number of objects, depth
and covered area, the only sample comparable to ours is the COMBO-17 sample 
\citep{Wolf2003}, which, in this redshift range, consists of 60 AGN candidates
over 0.78 square degree. At a similar depth, in terms of absolute magnitude,
we show also the data from the sample of \cite{Hunt2004}, which however consists
of 11 AGN in the redshift range 
$<z> \pm  \sigma_{z}=$3.03$\pm$0.35 \citep{Steidel2002}. Given the small number
of objects, the corresponding Hunt model fit was derived including also the
Warren data points \citep{Warren1994}. Moreover, they assumed the \cite{Pei1995} luminosity
evolution model, adopting the same values for $L^*$ and $\Phi^*$,
leaving free to vary the two slopes, both at the faint and at the bright end of
the LF. For comparison we show also the original Pei model fit derived from the empirical luminosity function estimated 
by \cite{Hartwick1990} and \cite{Warren1994}. 
In the same plot we show also the model fit derived from a sample of $\sim$100 
$z \sim 3$ (U-dropout) QSO candidates by Siana et al. (private comunication; 
see also \citealt{Siana2006}). 
This sample has been selected by using a simple optical/IR photometric selection 
at 19$< r' <$22 and the model fit has been derived by fixing the bright end slope 
at z=-2.85 as determined by SDSS data \citep{Richards2006QLF}.

In general, the comparison of the VVDS data points with those from the other
surveys shown in Figure \ref{fig:highz} shows a satisfactory agreement in the
region of overlapping magnitudes. 
The best model fit which reproduce our LF data points at $z\sim3$ is the Siana model 
with a faint end slope $\beta=-1.45$.
It is interesting to note that, in the faint
part of the LF, our data points appear to be higher with respect to the 
\cite{Hunt2004} fit and are instead closer to the extrapolation of the original
Pei model fit. This difference with the \cite{Hunt2004} fit is probably due to
the fact that, having only 11 AGN in their faint sample, their best fit to the
faint-end slope was poorly constrained.

\section{The bolometric luminosity function} \label{sez:BLF}

The comparison between the AGN LFs derived from samples selected in different
bands has been for a long time a critical point in the studies of the AGN
luminosity function. Recently, \cite{Hopkins2007}, combining a large number of
LF measurements obtained in different redshift ranges, observed wavelength bands
and luminosity intervals, derived the Bolometric QSO Luminosity Function in
the redshift range z = 0 - 6. For each observational band, they derived
appropriate bolometric corrections, taking into account the variation with
luminosity of both the average absorption properties (e.g. the QSO column
density N$_H$ from X-ray data) and the average global spectral energy
distributions. They show that, with these bolometric corrections, it is possible
to find a good agreement between results from all different sets of data.
 
\begin{figure}
\begin{center}                               
\includegraphics[height=13cm,width=8cm]{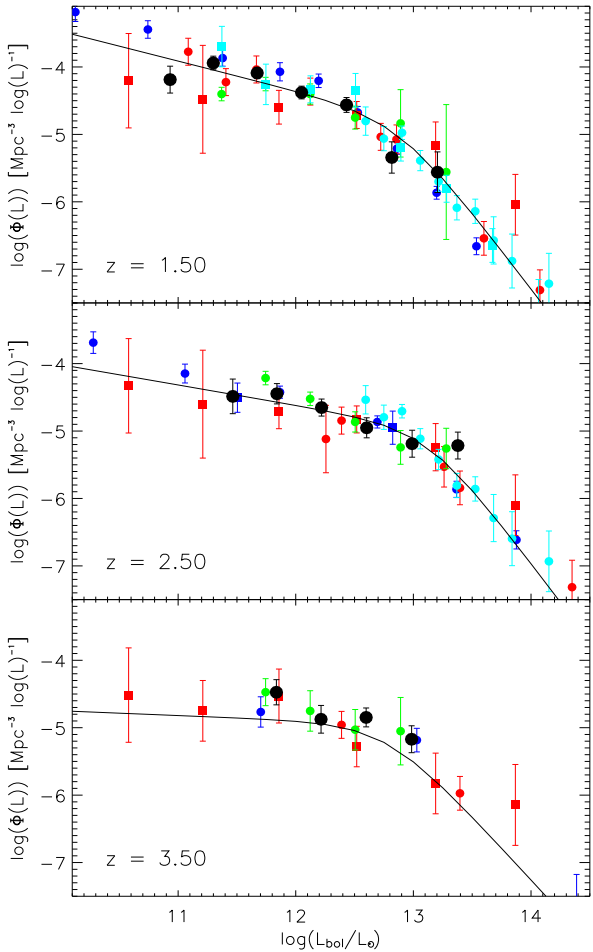}  
\caption{Bolometric luminosity function derived in three redshift bins
from our data (black dots), compared with \cite{Hopkins2007} best-fit model
and  the data-sets used in their work. The central redshift of
each bin is indicated in each panel. Here, we adopted the same color-code as in
\cite{Hopkins2007}, but for more clarity we limited the number of samples
presented in the Figure. Red symbols correspond to hard X-ray surveys
(squares: \citealt{Barger2005}; circles: \citealt{Ueda2003}).
Blue to soft X-ray surveys (squares: \citealt{Silverman2005b}; 
circles: \citealt{Hasinger2005}).
Cyan to infra-red surveys (circles: \citealt{Brown2006}; squares:
\citealt{Matute2006}). For the optical surveys we are showing here, with green
circles, the data from the COMBO-17 survey \citep{Wolf2003}, which is comparable
in depth to our sample.
}
\label{fig:BolLF}    
\end{center}
\end{figure}

We applied to our LF data points the bolometric corrections given by Eqs. (2)
and (4) of \cite{Hopkins2007} for the B-band and we derived the bolometric LF
shown as black dots in Figure \ref{fig:BolLF}. The solid line represents
the bolometric LF best fit model derived by \cite{Hopkins2007} and the colored
data points correspond to different samples: green points are from optical LFs,
blue and red points are from soft-X and hard-X LFs, respectively, and finally
the cyan points are from the mid-IR LFs. All these bolometric LFs data points
have been derived following the same procedure described in \cite{Hopkins2007}.

Our data, which sample the faint part of the bolometric luminosity function
better than all previous optically selected samples, are in good agreement
with all the other samples, selected in different bands.
Only in the last redshift bin, our data are quite higher with respect to 
the samples selected in other wavelength bands. The agreement remains however good 
with the COMBO-17 sample which is the only optically selected sample plotted here. 
This effect can be attributed to the fact that the conversions used to compute the 
Bolometric LF, being derived expecially for AGN at low redshifts, 
become less accurate at high redshift.
 
Our data show moreover good agreement also with
the model fit derived by \cite{Hopkins2007}. By trying various analytic fits to
the bolometric luminosity function \cite{Hopkins2007} concluded that
neither pure luminosity nor pure density evolution represent well all the data.
An improved fit can instead be obtained with a luminosity dependent density 
evolution model (LDDE) or, even better, with a PLE model in which both the 
bright- and the faint-end slopes evolve with redshift. 
Both these models can reproduce the observed flattening with redshift of the 
faint end of the luminosity function.

\section{Model fitting} \label{sez:ModelFitting}

In this Section we discuss the results of a number of different fits to our data as
a function of luminosity and redshift. For this purpose, we computed the luminosity function 
in 5 redshift bins at 1.0 $< z <$ 4.0 where the VVDS AGN sample consists of 121 objects. 
Since, in this redshift range, our data cover only the faint part of
the luminosity function, these fits have been performed by combining our data with
the LF data points from the SDSS data release 3 (DR3) \citep{Richards2006QLF} in
the redshift range 0 $< z <$ 4. The advantage of using the SDSS sample,
rather than, for example, the 2dF sample, is that the former sample, because of the
way it is selected, probes the luminosity function to much higher redshifts.
The SDSS sample contains more than 15,000
spectroscopically confirmed AGN selected from an effective area of 1622 sq.deg.
Its limiting magnitude (i $<$ 19.1 for z $<$ 3.0 and i $<$ 20.2 for z $>$ 3.0)
is much brighter than the VVDS and because of this it does not sample well the
AGN in the faint part of the luminosity function. For this reason,
\cite{Richards2006QLF} fitted the SDSS data using only a single power law, which
is meant to describe the luminosity function above the break luminosity. Adding
the VVDS data, which instead mainly sample the faint end of the luminosity
function, and analyzing the two samples together, allows us to cover the entire
luminosity range in the common redshift range (1.0 $< z <$ 4.0), also extending 
the analysis at z $<$ 1.0 where only SDSS data are available.

The goodness of fit between the computed LF data points and the various models is
then determined by the $\chi^2$ test.

For all the analyzed models we have parameterized the luminosity function as a
double power law that, expressed in luminosity, is given by:
\begin{equation}
\Phi(L,z) = \frac{\Phi_L^*}{(L/L^*)^{-\alpha} + (L/L^*)^{-\beta}}
\label{eq:phil}
\end{equation}
where $\Phi^*_L$ is the number of AGN per $Mpc^3$, L$^*$  is the characteristic
luminosity around which the slope of the luminosity function is changing and 
$\alpha$ and $\beta$ are the two power law indices. 
Equation \ref{eq:phil} can be expressed in absolute magnitude \footnote{
$\Phi_M^*=\Phi_L^* L^* \cdot \left|{ln 10^{-0.4}}\right|$} as:

\begin{equation}
\Phi(M,z) = \frac{\Phi_M^*}{10^{0.4(\alpha+1)(M-M^*)}+10^{0.4(\beta+1)(M-M^*)}} 
\label{eq:phim}
\end{equation}

\subsection{The PLE and PDE models}

\begin{figure*}
\begin{center}
\includegraphics[height=18cm,width=18cm]{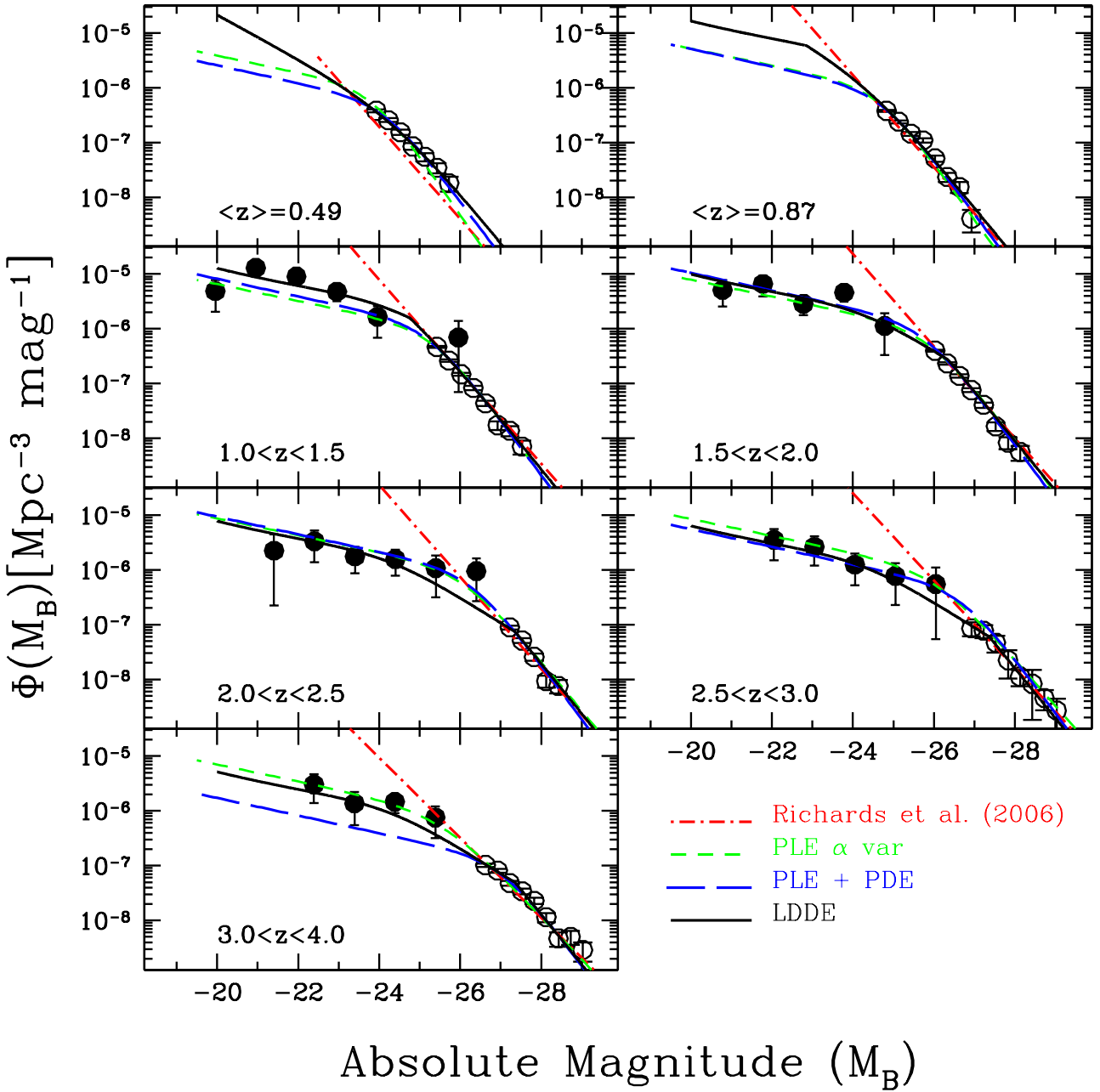}
\caption{Filled circles correspond to our rest-frame B-band luminosity function data points, derived 
in the redshift bins $1.0<z<1.5$, $1.5<z<2.0$, $2.0<z<2.5$, $2.5<z<3.0$ and
$3.0<z<4.0$. Open circles are the data points from the SDSS Data Release 3
(DR3) by \cite{Richards2006QLF}. These data are shown also in two redshift bins
below z = 1. The red dot-dashed line corresponds to the
model fit derived by \cite{Richards2006QLF} only for the SDSS data.
The other lines correspond to model fits derived considering the combination
of the VVDS and SDSS samples for different evolutionary models, as listed in
Table \ref{tab:fit} and described in Section \ref{sez:ModelFitting}.}	
\label{fig:LF_fit_a}
\end{center}
\end{figure*}

The first model that we tested is a Pure Luminosity Evolution (PLE) with the
dependence of the characteristic luminosity described by a 2nd-order polynomial
in redshift:

\begin{equation}
M^*(z) = M^*(0) - 2.5(k_1z + k_2z^2). 
\label{eq:PLE}
\end{equation}
Following the finding by \cite{Richards2006QLF} for the SDSS sample, we have
allowed a change (flattening with redshift) of the bright end slope according
to a linear evolution in redshift: $\alpha(z) =\alpha(0) + A~z$.  
The resulting best fit parameters are listed in the first line of Table 
\ref{tab:fit} and the resulting model fit is shown as a green short dashed line
in Figure \ref{fig:LF_fit_a}. The bright end slope $\alpha$ derived by our fit
($\alpha_{\rm VVDS}$=-3.19 at z=2.45) is consistent with the one found
by \cite{Richards2006QLF} ($\alpha_{\rm SDSS}$ = -3.1). 
\footnote{in their parameterization A$_1$=-0.4($\alpha+1)=$0.84}

This model, as shown in Figure \ref{fig:LF_fit_a}, while reproduces well the
bright part of the LF in the entire redshift range, does not fit the faint part
of the LF at low redshift (1.0 $< z <$ 1.5). This appears to be due to the fact
that, given the overall best fit normalization, the derived faint end slope 
($\beta=$-1.38) is too shallow to reproduce the VVDS data in this redshift
range.

\cite{Richards2005}, working on a combined 2dF-SDSS (2SLAQ) sample of AGN up to
$z=2.1$. found that, fixing all of the parameters except $\beta$ and the
normalization, to those of \cite{Croom2004}, the resulting faint end slope is
$\beta =-1.45 \pm 0.03$. This value would describe better our faint LF at low
redshift.
This trend suggests a kind of combined luminosity and density evolution not
taken into account by the used model. For this reason, we attempted to fit the
data also including a term of density evolution in the form of:
\begin{equation}
\Phi_M^*(z)=\Phi_M^*(0) \cdot 10^{k_{1D}z+k_{2D}z^2}  \label{eq:PLE1}
\end{equation}

In this model the evolution of the LF is described by both a term of
luminosity evolution, which affects $M^*$, and a term of density evolution,
which allows for a change in the global normalization $\Phi^*$. 
The derived best fit parameters of this model are listed in the second line of
Table \ref{tab:fit} and the model fit is shown as a blue long dashed line in
Figure \ref{fig:LF_fit_a}. This model gives a better $\chi^2$ with respect to
the previous model, describing the entire sample better than a simple PLE (the
reduced  $\chi^2$ decreases from $\sim$ 1.9 to $\sim$ 1.35).
However, it still does not satisfactorily reproduce the excess of faint objects
in the redshift bin 1.0 $< z <$ 1.5 and, moreover, it underestimates the faint
end of the LF in the last redshift bin (3.0 $< z <$ 4.0).

\subsection{The LDDE model}

Recently, a growing number of observations at different redshifts, in soft and
hard X-ray bands, have found evidences of a flattening of the faint end slope of
the LF towards high redshift. This trend has been described through a 
luminosity-dependent density evolution parameterization. Such a parameterization
allows the redshift of the AGN density peak to change as a function of
luminosity. This could help in explaining the excess of faint AGN found in the
VVDS sample at 1.0 $< z <$ 1.5. Therefore, we considered a luminosity dependent
density evolution model (LDDE), as computed in the major X-surveys 
(\citealt{Miyaji2000}; \citealt{Ueda2003}; \citealt{Hasinger2005}).
In particular, following \cite{Hasinger2005}, we assumed an LDDE evolution of 
the form:

\begin{equation}
\Phi(M_{\rm B},z)=\Phi(M,0)*e_{\rm d}(z,M_{\rm B})
\end{equation}
where:

\begin{equation}
e_{\rm d}(z,M_{\rm B}) = \left\{ 
	\begin{array}{ll}
	(1+z)^{p1} & (z \leq z_{\rm c}) \\ 
        e_{\rm d}(z_{\rm c})[(1+z)/(1+z_{\rm c})]^{p2} & (z > z_{\rm c})\\
	\end{array}
       \right. .
\label{eq:ldde1}
\end{equation}
along with

\begin{equation}
z_{\rm c}(M_{\rm B}) = \left\{ 
	\begin{array}{ll}
	z_{\rm c,0}10^{-0.4\gamma(M_{\rm B}-M_{\rm c})}& 
	(M_{\rm B} \geq M_{\rm c}) \\ 
        z_{\rm c,0} & (M_{\rm B} < M_{\rm c})\\
	\end{array}
       \right. .
\label{eq:ldde2}
\end{equation}
where z${\rm _c}$ corresponds to the redshift at which the evolution changes.
Note that z${\rm _c}$ is not constant but it depends on the luminosity.
This dependence 
allows different evolutions at different luminosities and can indeed reproduce
the differential AGN evolution as a function of luminosity, thus modifying the 
shape of the luminosity function as a function of redshift.  
We also considered two different assumptions for p1 and p2: (i) both parameters
constant and (ii) both linearly depending on luminosity as follows:

\begin{eqnarray}
p1 (M_{\rm B})= p1_{M_{\rm ref}} - 0.4 \epsilon_1\;(M_{\rm B}-M_{\rm ref}) \\ 
p2 (M_{\rm B})= p2_{M_{\rm ref}} - 0.4 \epsilon_2\;(M_{\rm B}-M_{\rm ref})
\end{eqnarray}

The corresponding $\chi^2$ values for the two above cases are respectively 
$\chi^2$=64.6 and $\chi^2$=56.8. Given the relatively small improvement of the
fit, we considered the addition of the two further parameters 
($\epsilon_1$ and $\epsilon_2$) unnecessary.
The model with constant p1 and p2 values is shown with a solid black line in
Figure \ref{fig:LF_fit_a} and the best fit parameters derived for this model
are reported in the last line of Table \ref{tab:fit}.

This model reproduces well the overall shape of the luminosity function over
the entire redshift range, including the excess of faint AGN at 1.0 $< z <$ 1.5.
The $\chi^2$ value for the LDDE model is in fact the best among all the 
analyzed models. We found in fact a $\chi^2$ of 64.6 for 67 degree of 
freedom and, as the reduced $\chi^2$ is below 1, it is acceptable 
\footnote{We note that the reduced $\chi^2$ of our best
fit model, which includes also VVDS data, is significantly better than that
obtained by \cite{Richards2006QLF} in fitting only the SDSS DR3 data.}. 

The best fit value of the faint end slope, which in this model corresponds to
the slope at z = 0, is $\beta$ =-2.0. This value is consistent with that derived
by \cite{Hao2005} studying the emission line luminosity function of a sample of
Seyfert galaxies at very low redshift (0 $< z <$ 0.15), extracted from the SDSS.
They in fact derived a slope $\beta$ ranging from -2.07 to -2.03, depending on
the line (H$\alpha$, [O\,\textsc{ii}] or [O\,\textsc{iii}]) used to compute the nuclear luminosity.
Moreover, also the normalizations are in good agreement, confirming our model also in 
a redshift range where data are not available and indeed leading us to have a good 
confidence on the extrapolation of the derived model.

\begin{table*}
\begin{center}
\begin{tabular}{c c c c c c|| c c c c c c}
\hline
\noalign{\smallskip}
\multicolumn{2}{c}{}&\multicolumn{4}{c||}{$1.0 < z < 1.5$}&\multicolumn{2}{c}{}&\multicolumn{4}{c}{$1.5 < z < 2.0$}\\
\noalign{\smallskip}
\hline
\noalign{\smallskip}
\multicolumn{2}{c}{$\Delta$M}&N$_{qso}$&Log$\Phi$(B)&\multicolumn{2}{c||}{$\Delta$Log$\Phi$(B)}&\multicolumn{2}{c}{$\Delta$M}&N$_{qso}$&Log$\Phi$(B)&\multicolumn{2}{c}{$\Delta$Log$\Phi$(B)}\\
\noalign{\smallskip}
\hline
\noalign{\smallskip}
-19.46 &  -20.46  &   3 &  -5.31  &  +0.20   &	  -0.38  &          &          &      &        &        &         \\
-20.46 &  -21.46  &  11 &  -4.89  &  +0.12   &	  -0.16  &   -20.28 &  -21.28  &   4  & -5.29  &  +0.18  &  -0.30 \\
-21.46 &  -22.46  &  17 &  -5.04  &  +0.09   &	  -0.12	&   -21.28 &  -22.28  &   7  & -5.18  &  +0.15  &  -0.22 \\
-22.46 &  -23.46  &   9 &  -5.32  &  +0.13   &	  -0.18	&   -22.28 &  -23.28  &   7  & -5.54  &  +0.14  &  -0.20 \\
-23.46 &  -24.46  &   3 &  -5.78  &  +0.20   &	  -0.38	&   -23.28 &  -24.28  &  10  & -5.34  &  +0.12  &  -0.17 \\
-25.46 &  -26.46  &   1 &  -6.16  &  +0.52   &	  -0.76	&   -24.28 &  -25.28  &   2  & -5.94  &  +0.23  &  -0.53 \\
\hline
\hline
\noalign{\smallskip}
\multicolumn{2}{c}{}&\multicolumn{4}{c||}{$2.0 < z < 2.5$}&\multicolumn{2}{c}{}&\multicolumn{4}{c}{$2.5 < z < 3.0$}\\
\noalign{\smallskip}
\hline
\noalign{\smallskip}
\multicolumn{2}{c}{$\Delta$M}&N$_{qso}$&Log$\Phi$(B)&\multicolumn{2}{c||}{$\Delta$Log$\Phi$(B)}&\multicolumn{2}{c}{$\Delta$M}&N$_{qso}$&Log$\Phi$(B)&\multicolumn{2}{c}{$\Delta$Log$\Phi$(B)}\\
\noalign{\smallskip}
\hline
\noalign{\smallskip}
  -20.90 &  -21.90    & 1 &  -5.65  &  +0.52  &  -0.76&	      & 	 &	&	  &	      &       \\ 
  -21.90 &  -22.90    & 3 &  -5.48  &  +0.20  &  -0.38&  -21.55  & -22.55  &   3  & -5.45 &   +0.20  &  -0.38 \\ 
  -22.90 &  -23.90    & 4 &  -5.76  &  +0.18  &  -0.30&  -22.55  & -23.55  &   4  & -5.58 &   +0.19  &  -0.34 \\ 
  -23.90 &  -24.90    & 4 &  -5.81  &  +0.18  &  -0.30&  -23.55  & -24.55  &   3  & -5.90 &   +0.20  &  -0.38 \\ 
  -24.90 &  -25.90    & 2 &  -5.97  &  +0.23  &  -0.53&  -24.55  & -25.55  &   2  & -6.11 &   +0.23  &  -0.53 \\ 
  -25.90 &  -26.90    & 2 &  -6.03  &  +0.23  &  -0.55&  -25.55  & -26.55  &   1  & -6.26 &   +0.52  &  -0.76 \\    
\hline
\hline
\noalign{\smallskip}
\multicolumn{2}{c}{}&\multicolumn{4}{c||}{$3.0 < z < 4.0$}&\multicolumn{2}{c}{}&\multicolumn{4}{c}{}\\
\noalign{\smallskip}
\hline
\noalign{\smallskip}
\multicolumn{2}{c}{$\Delta$M}&N$_{qso}$&Log$\Phi$(B)&\multicolumn{2}{c||}{$\Delta$Log$\Phi$(B)}&\multicolumn{6}{c}{}\\
\noalign{\smallskip}
\hline
\noalign{\smallskip}
  -21.89 &  -22.89    & 4  & -5.52 &   +0.19 &   -0.34  &\multicolumn{6}{c}{}  \\     
  -22.89 &  -23.89    & 3  & -5.86 &   +0.20 &   -0.40  &\multicolumn{6}{c}{} \\ 
  -23.89 &  -24.89    & 7  & -5.83 &   +0.14 &   -0.21  &\multicolumn{6}{c}{} \\ 
  -24.89 &  -25.89    & 3  & -6.12 &   +0.20 &   -0.38  &\multicolumn{6}{c}{} \\  
\hline
\hline
\end{tabular}
\end{center}
\caption{Binned luminosity function estimate for $\Omega_m$=0.3, $\Omega_{\Lambda}$=0.7 and 
H$_0$=70 km $\cdot$ s$^{-1}$ $\cdot$ Mpc$^{-1}$.
We list the values of Log $\Phi$ and the corresponding 1$\sigma$ errors in five redshift ranges, 
as plotted with full circles in Figure \ref{fig:LF_fit_a} and in $\Delta$M$_{B}$=1.0
magnitude bins. We also list the number of AGN contributing to the luminosity function estimate
in each bin } 
\label{tab:lf}
\end{table*}

\begin{table*}
\begin{center}
\begin{tabular}{c c c c c c c c c c c c}
\hline
\hline
Sample - Evolution Model  & $\alpha$ &$\beta$&$M^*$&$k_{1L}$ &$k_{2L}$ & $A$ & $k_{1D}$ &$k_{2D}$ &  $\Phi^*$&$\chi^2$&$\nu$\\
\hline
VVDS+SDSS - PLE $\alpha$ var &-3.83 &-1.38&-22.51 &1.23&-0.26&0.26&-&-&9.78E-7&130.36&69\\
VVDS+SDSS - PLE+PDE &-3.49 &-1.40&-23.40 &0.68&-0.073 &-&-0.97&-0.31&2.15E-7&91.4&68\\
\hline
Sample - Evolution Model  & $\alpha$ & $\beta$ & $M^*$ & $p1$ & $p2$ & $\gamma$ & $z_{\rm c,0}$ & $M_{\rm c}$ & $\Phi^*$ & $\chi^2$ & $\nu$\\
\hline
VVDS+SDSS - LDDE&-3.29&-2.0&-24.38&6.54&-1.37&0.21&2.08&-27.36&2.79E-8&64.6&67\\
\hline
\hline
\end{tabular}
\end{center}
\caption{Best fit models derived from the $\chi^2$ analysis of the combined sample VVDS+SDSS-DR3 in the redshift range $0.0< z
<4.0$ assuming a flat ($\Omega_m + \Omega_{\Lambda} = 1$) universe with $\Omega_m = 0.3$. 
}

\label{tab:fit}
\end{table*}

\section{The AGN activity as a function of redshift}

\begin{figure}
\begin{center}
\includegraphics[height=8cm,width=8cm]{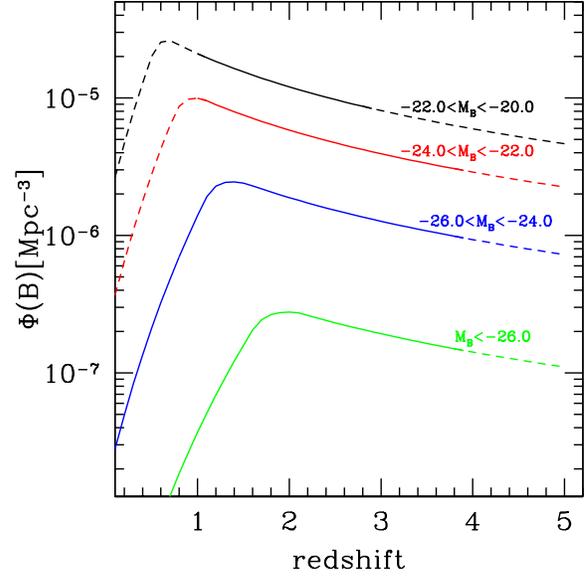}
\caption{Evolution of comoving AGN space density with redshift, for different luminosity range:
-22.0$< M_{B} <$-20.0; -24.0$< M_{B} <$-22.0; -26.0$< M_B <$-24.0 and M$_{B} <$-26.0. 
Dashed lines correspond to the redshift range in which the model has been extrapolated.}
\label{fig:peak}
\end{center}
\end{figure}

By integrating the luminosity function corresponding to our best fit model (i.e
the LDDE model; see Table \ref{tab:fit}), we derived the comoving 
AGN space density as a function of redshift for different luminosity ranges
(Figure \ref{fig:peak}). 

The existence of a peak at z$\sim$ 2 in the space density of bright AGN is known
since a long time, even if rarely it has been possible to precisely locate the
position of this maximum within a single optical survey.
Figure \ref{fig:peak} shows that for our best fit model the peak of the AGN
space density shifts significantly towards lower redshift going to lower
luminosity. The position of the maximum moves from z$\sim$ 2.0 for M$_B<$-26.0
to z$\sim$ 0.65 for -22$<M_B<$-20. 

A similar trend has recently been found by the analysis of several deep X-ray
selected samples \citep{Cowie2003, Hasinger2005,LaFranca2005}. 
To compare with X-ray results, by applying the same bolometric corrections used is 
Section \ref{sez:BLF}, 
we derived the volume densities derived by our best fit LDDE model in the same 
luminosity ranges as \cite{LaFranca2005}.
We found that the volume density peaks at z $\simeq$ [0.35; 0.7; 1.1; 1.5] respectively for
LogL$_{X(2-10 kev)}$ = [42--43; 43--44; 44--44.5; 44.5--45]. In the same luminosity 
intervals, the values for the redshift of the peak obtained by \cite{LaFranca2005} are 
z $\simeq$ [0.5; 0.8; 1.1; 1.5], in good agreement with our result.
This trend has been interpreted as
evidence of AGN (i.e. black hole) ``cosmic downsizing'', similar to what has
recently been observed in the galaxy spheroid population \citep{Cimatti2006}.
The downsizing \citep{Cowie1996} is a term which is used to describe the
phenomenon whereby luminous activity (star formation and accretion onto black
holes) appears to be occurring predominantly in progressively lower mass
objects (galaxies or BHs) as the redshift decreases. As such, it explains 
why the number of bright sources peaks at higher redshift than the number of
faint sources.

As already said, this effect had not been seen so far in the analysis of
optically selected samples. This can be due to the fact that most of the optical
samples, because of their limiting magnitudes, do not reach luminosities where
the difference in the location of the peak becomes evident. 
The COMBO-17 sample \citep{Wolf2003}, for example, even if it covers enough 
redshift range ($1.2<z<4.8$) to enclose the peak of the AGN activity, 
does not probe luminosities faint enough to find a significant indication for
a difference between the space density peaks of AGN of different luminosities
(see, for example, Figure 11 in \cite{Wolf2003}, which is analogous to our Figure
\ref{fig:peak}, but in which only AGN brighter than M $\sim$ -24 are shown). 
The VVDS sample, being about one magnitude fainter than the COMBO-17 sample and
not having any bias in finding faint AGN, allows us to detect for the first time
in an optically selected sample the shift of the maximum space density towards
lower redshift for low luminosity AGN.

\section{Summary and conclusion}

In the present paper we have used the new sample of AGN, collected by the VVDS
and presented in \cite{Gavignaud2006}, to derive the optical luminosity function
of faint type--1 AGN.

The sample consists of 130 broad line AGN (BLAGN) selected on the basis of only 
their spectral features, with no morphological and/or color selection biases.
The absence of these biases is particularly important for this sample because
the typical non-thermal AGN continuum can be significantly masked by the
emission of the host galaxy at the low intrinsic luminosity of the VVDS AGN. 
This makes the optical selection of the faint AGN candidates very difficult
using the standard color and morphological criteria. Only spectroscopic surveys
without any pre-selection can therefore be considered complete in this
luminosity range.

Because of the absence of morphological and color selection, our sample shows
redder colors than those expected, for example, on the basis of the color track
derived from the SDSS composite spectrum and the difference is stronger for
the intrinsically faintest objects. Thanks to the extended multi-wavelength
coverage in the deep VVDS fields in which we have, in addition to the optical
VVDS bands, also photometric data from GALEX, CFHTLS, UKIDSS and SWIRE, we
examined the spectral energy distribution of each object and we fitted it with
a combination of AGN and galaxy emission, allowing also for the possibility 
of extinction of the AGN flux. We found that both effects (presence of dust
and contamination from the host galaxy) are likely to be responsible for this
reddening, even if it is not possible to exclude that faint AGN are
intrinsically redder than the brighter ones. 

We derived the luminosity function in the B-band for $1 < z < 3.6$, using the
usual $1/V_{max}$ estimator \citep{Schmidt1968}, which gives the space density
contributions of individual objects. Moreover, using the prescriptions recently
derived by \cite{Hopkins2007}, we computed also the bolometric luminosity 
function for our sample. This allows us to compare our results also with other
samples selected from different bands. 

Our data, more than one magnitude fainter than previous optical surveys, 
allow us to constrain the faint part of the luminosity function up to high
redshift. A comparison of our data with the 2dF sample at low redshift (1 $<$ z
$<$ 2.1) shows that the VVDS data can not be well fitted with the PLE models
derived by previous samples. Qualitatively, our data suggest the presence 
of an excess of faint objects at low redshift ($1.0<z<1.5$) with respect to these models.

Recently, a growing number of observations at different redshifts, in soft and 
hard X-ray bands, have found in fact evidences of a similar trend and they have
been reproduced with a luminosity-dependent density evolution parameterization.
Such a parameterization allows the redshift of the AGN density peak to change
as a function of luminosity and explains the excess of faint AGN that we found
at 1.0 $< z <$ 1.5. Indeed, by combining our faint VVDS sample with the large
sample of bright AGN extracted from the SDSS DR3 \citep{Richards2006QLF},
we found that the evolutionary model which better represents the combined
luminosity functions, over a wide range of redshift and luminosity, is an 
LDDE model, similar to those derived from the major X-surveys. The derived
faint end slope at z=0 is $\beta$ = -2.0, consistent with the value derived by 
\cite{Hao2005} studying the emission line luminosity function of a sample of
Seyfert galaxies at very low redshift.

A feature intrinsic to these LDDE models is that the comoving AGN space density
shows a shift of the peak with luminosity, in the sense that more luminous AGN
peak earlier in the history of the Universe (i.e. at higher redshift), while the
density of low luminosity ones reaches its maximum later (i.e. at lower
redshift). In particular, in our best fit LDDE model the peak of the space
density ranges from z $\sim$ 2 for M$_B <$ -26 to z$\sim$ 0.65 for 
-22 $< M_B <$ -20. This effect had not been seen so far in the analysis of
optically selected samples, probably because most of the optical samples do
not sample in a complete way the faintest luminosities, where the difference in
the location of the peak becomes evident. 

Although the results here presented appear to be already robust, the larger AGN
sample we will have at the end of the still on-going VVDS survey ($>$ 300 AGN),
will allow a better statistical analysis and a better estimate of the parameters
of the evolutionary model.

\begin{acknowledgements}
This research has been developed within the framework of the VVDS
consortium.\\
This work has been partially supported by the
CNRS-INSU and its Programme National de Cosmologie (France),
and by Italian Ministry (MIUR) grants
COFIN2000 (MM02037133) and COFIN2003 (num.2003020150).\\
Based on data obtained with the European Southern Observatory
Very Large Telescope, Paranal, Chile, program 070.A-9007(A),
272.A-5047, 076.A-0808, and on data obtained at the Canada-France-Hawaii
Telescope, operated by the CNRS of France, CNRC in Canada, and the
University of Hawaii.
The VLT-VIMOS observations have been carried out on guaranteed
time (GTO) allocated by the European Southern Observatory (ESO)
to the VIRMOS consortium, under a contractual agreement between the
Centre National de la Recherche Scientifique of France, heading
a consortium of French and Italian institutes, and ESO,
to design, manufacture and test the VIMOS instrument.
\end{acknowledgements}

\end{document}